# Superconductivity and its Enhancement under High Pressure in "F-free" Single Crystals of CeOBiS$_2$

Masashi Tanaka[a,b,*,#], Masanori Nagao[c,*,#], Ryo Matsumoto[b,d], Noriyuki Kataoka[e], Ikuo Ueta[c],

Hiromi Tanaka[e], Satoshi Watauchi[c], Isao Tanaka[c], and Yoshihiko Takano[b,d]

[a]Graduate School of Engineering, Kyushu Institute of Technology, 1-1 Sensui-cho, Tobata, Kitakyushu

804-8550, Japan

[b]WPI-MANA, National Institute for Materials Science, 1-2-1 Sengen, Tsukuba, Ibaraki 305-0047, Japan

[c]University of Yamanashi, 7-32 Miyamae, Kofu, Yamanashi 400-8511, Japan

[d]University of Tsukuba, 1-1-1 Tennodai, Tsukuba, Ibaraki 305-8577, Japan

[e]National Institute of Technology, Yonago College, 4448 Hikona, Yonago, Tottori 683-8502, Japan

**Corresponding Authors**

*Masashi Tanaka for the correspondence of crystallographic analysis and high pressure measurement
E-mail: mtanaka@mns.kyutech.ac.jp
Postal address: Graduate School of Engineering, Kyushu Institute of Technology,
1-1 Sensui-cho, Tobata, Kitakyushu 804-8550, Japan
Telephone/FAX number: (+81)93-884-3204

*Masanori Nagao for the correspondence of crystal growth and the compositional analysis
E-mail: mnagao@yamanashi.ac.jp
Postal address: University of Yamanashi, Center for Crystal Science and Technology
7-32 Miyamae, Kofu 400-8511, Japan
Telephone number: (+81)55-220-8610; Fax number: (+81)55-254-3035

**Abstract**

"F-free" CeOBiS$_2$ single crystals have successfully grown, thoroughly eliminating a concern about F-contamination by using a high-purity CsCl flux. The obtained crystals have a plate-like shape with a size of 1.0-3.0 mm in the well-developed plane. Single crystal X-ray structural analysis clearly revealed that the CeOBiS$_2$ crystallizes with a space group *P4/nmm* (with lattice parameters of $a$ = 4.0189(6) Å, $c$ = 13.573(2) Å). The bond valence sum estimation and X-ray photoelectron spectroscopy analysis showed that the chemical state of Ce was in the mixed valence of Ce$^{3+}$ and Ce$^{4+}$. The single crystals show superconductivity with zero resistivity at ~1.3 K. This is the first conclusive evidence of superconductivity driven by Ce valence fluctuation in surely non-doped CeOBiS$_2$. The superconducting transition temperature was enhanced up to ~3.8 K by applying hydrostatic pressure.

## 1. Introduction

In a development of materials science, it is important to prepare a series compound whose physical properties can be continuously controlled without changing its basic structure. Recently discovered BiS$_2$-based layered superconductors, $R$(O,F)BiS$_2$ ($R$: La, Ce, Pr, Nd, Yb) [1-5], are a kind of such compound family. These compounds are also categorized as "mixed-anion" materials, and are paid attention as a candidate for next generation functional materials, such as optoelectronic material [6], thermoelectric material [7,8], monolayer application [9], and so on. When the O atoms are substituted by F atoms, the parent material of $R$OBiS$_2$ continuously varies its physical properties, then it shows superconductivity. The compounds have layered structure which is composed of superconducting BiS$_2$ layers and charge reservoir $R$O layers, similar to which is found in the Fe-based superconductors, $Rn$(O,F)FeAs ($Rn$: rare earth elements) [10,11]. Many BiS$_2$-based superconducting materials have been developed by modifying the building blocks, for example, Bi(O,F)BiS$_2$ [12,13], La(O,F)Bi$Ch_2$ ($Ch$ = S, Se) [14,15].

Ce(O,F)BiS$_2$ is one of the material series in the BiS$_2$-based superconductors family. It has been reported to show curious phenomenon like a coexistence of magnetic ordering with superconductivity [16,17]. In order to understand the superconducting nature of this material series, it is important to evaluate the fundamental properties of parent compound, CeOBiS$_2$.

For such a purpose, a usage of the single crystals is definitely helpful. It brings us a lot of valuable information such as precise crystal structure [18-21], superconducting anisotropy [22], high pressure effect on uniaxial compression [23,24], and so on. The single crystals of BiS$_2$-based compounds can be prepared by using a conventional alkali metal chloride flux method [25-29], and CeOBiS$_2$ also has been prepared [30,31]. However, the reported results were controversial, an article showed that the $CeOBiS_2$ shows superconductivity without fluorine doping [30], on the other hand, the other article had reported nonmetallic behavior of resistivity in $CeOBiS_2$ [31].

The controversial results might be attributed to a small amount of fluorine contamination from the flux material. Because typical CsCl chemical reagent had been confirmed to contain F elements by ion chromatography analysis [30]. It is necessary to prepare $CeOBiS_2$ which is thoroughly eliminated the concern about F-contamination. In this study, we have successfully grown such "F-free" $CeOBiS_2$ single crystals by using a high-purity CsCl flux. The single crystals are characterized by means of X-ray structural analysis and X-ray photoelectron spectroscopy (XPS), then the superconductivity under high pressure as well as ambient pressure is discussed regarding the valence state of Ce. This is a part of studies to understand puzzles behind the superconductivity of the $BiS_2$-based compound family.

**2. Experimental**

*2.1 Preparation of single crystals*

Single crystals of $CeOBiS_2$ were grown by similar way to the previous studies by using an alkali metal flux method [19,26]. Powders of $Ce_2S_3$ (99.9 wt%), $Bi_2S_3$ (99.9 wt%), $Bi_2O_3$ (99.9 wt%) were mixed with the nominal composition of $CeOBiS_2$. The starting powders of 0.8 g were mixed with 5 g of high purity CsCl flux (Strem Chemicals, 99.999 wt%), and it was sealed in an evacuated quartz tube. The quartz tube was heated at 950°C for 10 h, followed by cooling to 650°C at a rate of 1°C/h, then the sample was cooled down to room temperature in the furnace. The obtained materials were washed and filtered by distilled water in order to remove the flux materials. In the case of "F-doped" crystals, BiF$_3$ powder was added to the starting materials as a fluorine source with a flux of CsCl/KCl [19].

*2.2 Characterization*

Single crystal X-ray structural analysis was carried out using a Rigaku Mercury CCD diffractometer with graphite monochromated MoKα radiation ($\lambda$ = 0.71072 Å) (Rigaku, XtaLABmini). The crystal structure was solved and refined by using the program SHELXT and SHELXL [32,33], respectively, in the WinGX software package [34]. The compositional ratio of the single crystals was evaluated by electron probe microanalysis (EPMA) associated with the observation of the microstructure by using scanning electron microscopy (SEM) (JEOL, JXA-8200).

Chemical states of the component were estimated by XPS analysis using AXIS-ULTRA DLD (Shimadzu/Kratos) with AlKα X-ray radiation ($h\nu$ = 1486.6 eV), operating under a pressure of the order of $10^{-9}$ Torr. The sample single crystals were cleaved before the measurement. The analyzed area was approximately 1×1 mm$^2$. The background signals were subtracted by using an active Shirley method on COMPRO software [35]. The photoelectron peaks were analyzed by the pseudo-Voigt functions peak fitting.

The temperature dependence of resistivity at ambient pressure was measured by physical property measurement system (PPMS, Quantum Design) with an adiabatic demagnetization refrigerator (ADR) option. The temperature dependence of magnetization was measured by a superconducting quantum interference device (SQUID) magnetometer (MPMS, Quantum Design) under zero-field cooling (ZFC) and field cooling (FC) with an applied field of $H$ = 10 Oe parallel to the *c*-axis of single crystals. Piston cylinder type high pressure cell was used in PPMS for measurement of resistivity under hydrostatic pressure by using Fluorinert 70/77 as a pressure medium. The pressure values were estimated from superconducting transition temperature ($T_c$) of Pb manometer.

## 3. Results and discussion

*3.1 Single crystal growth of "F-free" CeOBiS$_2$*

In an attempt to obtain the "F-free" CeOBiS$_2$ single crystals, a high purity CsCl was employed for the flux material. The inclusion of F atoms in the flux was confirmed to less than $6.53 \times 10^{-4}$ mol% by ion chromatography. Figure 1 shows a typical SEM image for CeOBiS$_2$ single crystals, demonstrating a plate-like shape with 1.0-3.0 mm in size and 10-30 μm in thickness. Ce, O, Bi, S elements were detected in this sample by qualitative analysis of EPMA. The averaged compositional ratio was estimated to be Ce:O:Bi:S = 1.00±0.10:1.06±0.10:1.01±0.05:2, in which the ratio was based on S = 2. It is in good agreement with the nominal composition within the error, suggesting that the single crystals have a composition of CeOBiS$_2$.

On the other hand, "F-doped" CeOBiS$_2$ crystals were also synthesized for a comparison in the similar synthesizing way [19]. The compositional ratio of the "F-doped" crystals was estimated to be Ce(O$_{0.73}$F$_{0.27}$)BiS$_2$ by EPMA. Hereafter, we call the "F-free" CeOBiS$_2$ and "F-doped" Ce(O$_{0.73}$F$_{0.27}$)BiS$_2$ crystals for CeOBiS$_2$ and Ce(O,F)BiS$_2$, respectively.

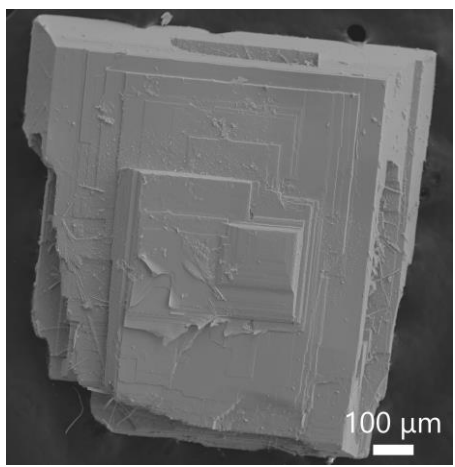

Figure 1. SEM image of CeOBiS$_2$ single crystals.

*3.2 Single crystal structural analysis of CeOBiS$_2$ and Ce(O,F)BiS$_2$*

Single crystal X-ray structural analyses of the CeOBiS$_2$ and Ce(O,F)BiS$_2$ were successfully performed. Details of the analysis and crystallographic parameters are listed in Table I and Table II. The refinement for CeOBiS$_2$ and Ce(O,F)BiS$_2$ was converged to the $R_1$ values of 6.47% and 7.79% for $I \geq 2\sigma(I)$, respectively, for 16 variables including anisotropic displacement parameters. Both compounds crystallize with space group *P4/nmm*, and they have lattice parameters of $a$ = 4.0189(6) Å, $c$ = 13.573(2) Å for CeOBiS$_2$, and $a$ = 4.0321(6) Å, $c$ = 13.356(3) Å for Ce(O,F)BiS$_2$. All atoms were located at special positions and their anisotropic displacement parameters were all positive with similar values, appearing physically reasonable. The refined occupancy of Bi is slightly smaller than 1.0, although those of the other elements were refined to be unity within the error. This fact may be attributed to existence of small deficiency in Bi site similar to that found in the other Bi*Ch*$_2$-based compound [18,20,36]. The final refinement in CeOBiS$_2$ was performed with fixed the occupancies of Ce, S, and O to 1.0. Since the occupancy refinement of O and F in Ce(O,F)BiS$_2$ did not converge under constrain of Occ.(O) + Occ.(F) = 1, the occupancies were fixed to the EPMA-estimated values of Occ.(O) = 0.73, Occ.(F) = 0.27 in the final refinement. It is necessary to mention about large residual peaks/holes around Bi and Ce sites. These large residual densities are attributed to disagreeable reflections due to the stacking fault, which frequently occurs in layer structured compounds, and/or the local disorder of Bi/Ce-sites in its crystal structure those found in the similar Bi$Ch_2$ compounds [20].

Table I. Atomic coordinates, atomic displacement parameters (Å$^2$), and bond valence sum (BVS) for the CeOBiS$_2$ and Ce(O,F)BiS$_2$.

| | Site | Wyck. | S.O.F | x/a | y/b | z/c | $U_{11}$ | $U_{22}$ | $U_{33}$ | $U_{eq}$ | BVS |
|---|---|---|---|---|---|---|---|---|---|---|---|
| CeOBiS$_2$ | | | | | | | | | | | |
| | Ce | 2c | 1.0 | 1/4 | 1/4 | 0.0925(2) | 0.0029(7) | 0.0029(7) | 0.0211(13) | 0.0089(6) | +3.19 |
| | Bi | 2c | 0.958(9) | 3/4 | 3/4 | 0.37319(10) | 0.0112(6) | 0.0112(6) | 0.0113(8) | 0.0112(5) | +2.96 |
| | S1 | 2c | 1.0 | 3/4 | 3/4 | 0.1880(6) | 0.010(3) | 0.010(3) | 0.009(4) | 0.009(2) | -2.10 |
| | S2 | 2c | 1.0 | 1/4 | 1/4 | 0.3783(7) | 0.010(3) | 0.010(3) | 0.030(5) | 0.017(2) | -1.84 |
| | O | 2a | 1.0 | 3/4 | 1/4 | 0 | 0.002(7) | 0.002(7) | 0.023(13) | 0.009(4) | -2.22 |
| Ce(O,F)BiS$_2$ | | | | | | | | | | | |
| | Ce | 2c | 1.0 | 1/4 | 1/4 | 0.0990(2) | 0.0084(8) | 0.0084(8) | 0.0164(14) | 0.01107(8) | +2.93 |
| | Bi | 2c | 0.962(11) | 3/4 | 3/4 | 0.37679(12) | 0.0107(7) | 0.0107(7) | 0.0128(9) | 0.0114(6) | +2.84 |
| | S1 | 2c | 1.0 | 3/4 | 3/4 | 0.1880(6) | 0.007(2) | 0.007(2) | 0.014(4) | 0.009(2) | -2.13 |
| | S2 | 2c | 1.0 | 1/4 | 1/4 | 0.3783(7) | 0.011(3) | 0.011(3) | 0.036(6) | 0.020(2) | -1.80 |
| | O/F | 2a | 0.73/0.27(Fix) | 3/4 | 1/4 | 0 | 0.010(6) | 0.010(6) | 0.013(12) | 0.011(5) | -1.84 |

Note: $U_{12}$, $U_{13}$ and $U_{23}$ are 0, and $U_{eq}$ is defined as one-third of the trace of the orthogonalized $U$ tensor

TABLE II. Crystallographic data for the CeOBiS$_2$ and Ce(O,F)BiS$_2$

| | | |
|---|---|---|
| Structural formula | Ce$_2$O$_2$Bi$_{1.92}$S$_4$ | Ce$_2$(O$_{1.46}$F$_{0.54}$)Bi$_{1.92}$S$_4$ |
| Formula weight | 840.89 | 844.18 |
| Crystal dimensions (mm) | 0.05 x 0.04 x 0.01 | 0.11 x 0.08 x 0.03 |
| Crystal shape | Platelet | Platelet |
| Crystal system | Tetragonal | Tetragonal |
| Space group | *P4/nmm* (No. 129) | *P4/nmm* (No.129) |
| $a$ (Å) | 4.0189(6) | 4.0321(6) |
| $c$ (Å) | 13.573(2) | 13.356(3) |
| $V$ (Å$^3$) | 219.23(7) | 217.14(8) |
| Z | 1 | 1 |
| $d_{calc}$ (g/cm$^3$) | 6.369 | 6.456 |
| Temperature (K) | 293 | 293 |
| $\lambda$ MoK$\alpha$ (Å) | 0.71073 | 0.71073 |
| $\mu$ (mm$^{-1}$) | 49.406 | 50.048 |
| Absorption correction | Empirical | Empirical |
| $\theta_{max}$ (°) | 32.581 | 32.642 |
| Index ranges | -5<h<6, -5<k<6, -20<l<19 | -5<h<6, -5<k<6, -20<l<19 |
| Total reflections | 2635 | 2547 |
| Unique reflections | 289 | 284 |
| Observed [$I \geq 2\sigma(I)$] | 207 | 256 |
| $R_{int}$ for all reflections | 0.270 | 0.329 |
| No. of variables | 16 | 16 |
| $R1/wR2$ [$I \geq 2\sigma(I)$] | 0.0647/0.1331 | 0.0779/0.2043 |
| $R1/wR2$ (all data) | 0.0862/0.1404 | 0.0817/0.2066 |
| GOF on $F_o^2$ | 0.879 | 1.063 |
| Max./Min. residual density (e$^-$/Å$^3$) | 4.03 / -8.53 | 12.28 / -5.52 |

A schematic illustration of the crystal structure of CeOBiS$_2$ is shown in Figure 2(a). The crystal has an alternate stacking of BiS$_2$ and CeO layers, isostructural with those reported in the other BiCh$_2$-based compounds [1,18,20,21,37]. Some selected bonding distances and angles are given in Fig. 2(b) as the thermal ellipsoids representation and Table III. Bi atoms are bonded with S atoms in three kinds such as Bi-S1, inter-plane Bi-S2, and intra-plane Bi-S2 bondings. The interatomic distance of intra-plane Bi-S2 is much shorter and longer than that of inter-plane Bi-S2 (3.373(10) Å) and Bi-S1 (2.514(8) Å), respectively, with similar tendency to the BiCh$_2$-based compounds [18-21]. The intra-plane S2-Bi-S2 angle in the CeOBiS$_2$ crystals is 177.204(3)°. The Bi-S plane is less distorted compare to the Bi-Ch plane in the other F-doped BiCh$_2$-based compounds [18,20,21]. As seen from Table III, the Bi-S plane in Ce(O,F)BiS$_2$ is 179.630(3)°, which is flatter than that of CeOBiS$_2$. The F-doping to the blocking layers in La(O,F)BiS$_2$ increases the flatness in Bi-S plane, then stabilizes the crystal structure [18]. The carrier numbers in the blocking layers may optimize the Bi-S plane. Namely, the corrugation adjustment in the Bi-S plane seems to be driven by the charge transfer between Bi-S plane and the blocking layers. In the case of this study, valence fluctuant Ce atoms would change its valence state with the structural optimization.

Bond valence sum's (BVS's) for CeOBiS$_2$ and Ce(O,F)BiS$_2$ were estimated from the observed bond distances and the BVS parameters ($r_0$ and $B$ values) of nominal valence Ce$^{3+}$, Bi$^{3+}$, S$^{2-}$, O$^{2-}$, F$^{1-}$ provided by Brown [38]. The BVS values are also listed in Table I. According to the refined composition of CeOBi$_{0.96}$S$_2$ in the "F-free" crystal, the nominal valence summation is -0.12. The charge neutrality should be kept by modification in the valence states of the cations, Ce and Bi. Note that the BVS of Bi was estimated to be nearly + 3, (+2.96). On the other hand, BVS of Ce was estimated to be +3.19, indicating the charge neutrality is conserved by the existence of mixed valence state of $Ce^{3+}$ and $Ce^{4+}$, although that of Ce(O,F)BiS$_2$ is +2.93 implying the nearly single valence state of $Ce^{3+}$.

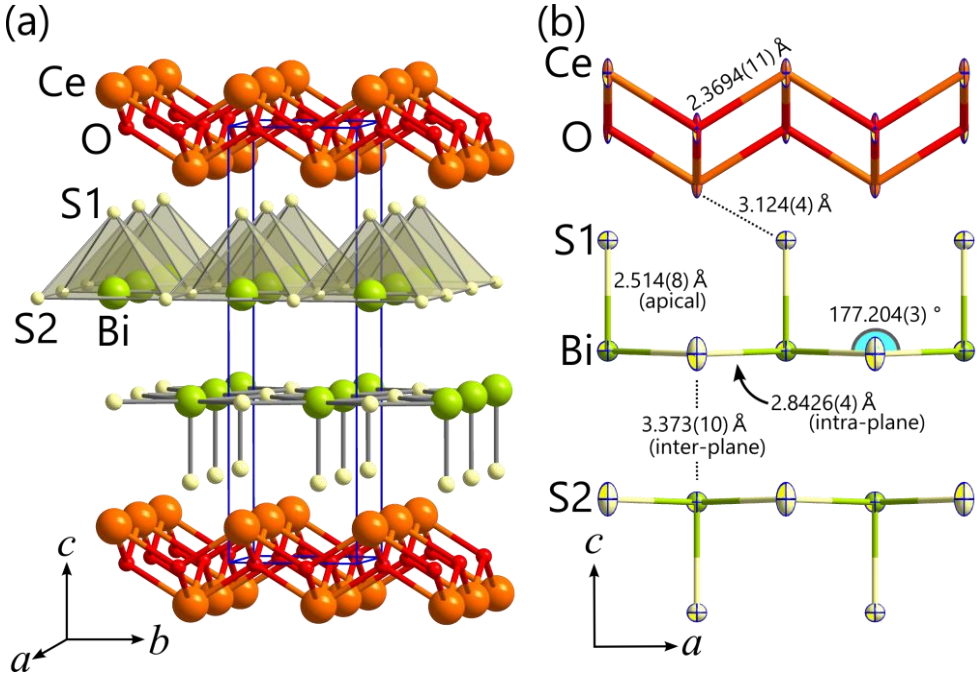

Figure 2. (a) Schematic illustration of the crystal structure of CeOBiS$_2$. (b) Thermal ellipsoids representation of CeOBiS$_2$ with the selected bonding distances and angles. Displacement ellipsoids are drawn at an 80% probability level.

Table III. Selected bond lengths (Å) and angles (°) of the CeOBiS$_2$ and Ce(O,F)BiS$_2$ single crystals

| Distance (Å) | CeOBiS$_2$ | Ce(O,F)BiS$_2$ |
| --- | --- | --- |
| Ce-O(F) × 4 | 2.3694(11) | 2.4107(14) |
| Ce-S1 × 4 | 3.124(4) | 3.082(4) |
| Bi-S1 | 2.514(8) | 2.540(10) |
| Bi-S2 [inter-plane] | 3.373(10) | 3.300(14) |
| Bi-S2 [intra-plane] × 4 | 2.8426(4) | 2.8511(3) |
| Angle (°) | | |
| Ce-O(F)-Ce | 116.006(3) | 113.503(3) |
| Ce-S1-Ce | 80.083(2) | 81.707(2) |
| S1-Bi-S2 | 91.40(5) | 89.8(2) |
| S2-Bi-S2 [inter-plane] | 89.966(2) | 90.2(2) |
| S2-Bi-S2 [intra-plane] | 177.204(3) | 179.630(3) |

*3.2 Valence states of Ce in CeOBiS$_2$ and Ce(O,F)BiS$_2$ single crystals*

XPS analysis was carried out to investigate the chemical states of Ce in the both single crystals of CeOBiS$_2$ and Ce(O,F)BiS$_2$. The XPS spectrum of the Ce 3$d$ in CeOBiS$_2$ is shown in the lower line of Figure 3. It is clearly seen that there is a single peak at 916.6 eV (u'''). It has been reported that it is associated to the Ce 3$d_{3/2}$ and a characteristic feature of a presence of tetravalent Ce ions (Ce$^{4+}$) in Ce compounds [39], implying that the chemical state of Ce atoms in the single crystals is in a mixed valence state of Ce$^{4+}$ and Ce$^{3+}$. Since the signal of the Ce 3$d$ level has a very complicated satellite structure, the peaks were labeled v and u according to the method firstly established by Burroughs et al [40]. The two multiplets v and u corresponds to the spin-orbit splitting of 3$d_{5/2}$ and 3$d_{3/2}$ core holes, respectively. In this notation, the corresponding spin-orbit splitting is indicated by the same superscript of v. The peak positions of Ce$^{4+}$ and Ce$^{3+}$ were firstly determined from measurements for standard materials of CeO$_2$ and Ce$_2$S$_3$, respectively. The v series of peaks correspond to a mixing configuration of v: Ce(IV)(3$d^9 4f^2$) O(2$p^4$), v'': Ce(IV)(3$d^9 4f^1$) O(2$p^5$) and v''': Ce(IV)(3$d^9 4f^0$) O(2$p^6$) final states in Ce$^{4+}$, and v$^0$: Ce(III)(3$d^9 4f^2$) O(2$p^5$) and v': Ce(III)(4$d^9 4f^1$) O(2$p^6$) final states in Ce$^{3+}$. The same assignments could be also applied to the u series of peaks. Therefore, ten peaks in total corresponding to the pairs of spin-orbit doublets can be identified in the Ce 3$d_{5/2,3/2}$ spectrum, namely the spectrum is composed of the peaks of Ce(III) (v$^0$ + v' + u$^0$ + u') and Ce(IV) (v + v'' + v''' + u + u'' + u'''). These results are in good agreement with the other XPS results in mixed valence Ce compounds [39,41]. The area ratio of the fitted peaks of Ce$^{3+}$ to that of Ce$^{4+}$ was estimated to be about 25:6, suggesting the average valence of Ce is Ce$^{3.19+}$. This value is identical to the result of BVS estimation from the X-ray structural analysis.

The XPS result of Ce(O,F)BiS$_2$ is also shown in the upper line of Fig. 3. Note that the u''' peak, which is a fingerprint of Ce$^{4+}$, hardly observed in the spectrum. The Ce$^{4+}$ valence state seems to disappear after the F-doping. The average Ce valence was estimated to be Ce$^{3.04+}$ from this measurement. This valence number is also in good agreement with the BVS estimation, and it indicates that the Ce valence no longer fluctuates in the F-doped crystals. These valence states may change the orbital overlapping in the BiS$_2$ plane due to the corrugation adjustment, and affect to the physical properties taking into account the in-plane chemical pressure effect of Bi$Ch_2$-based compounds [42].

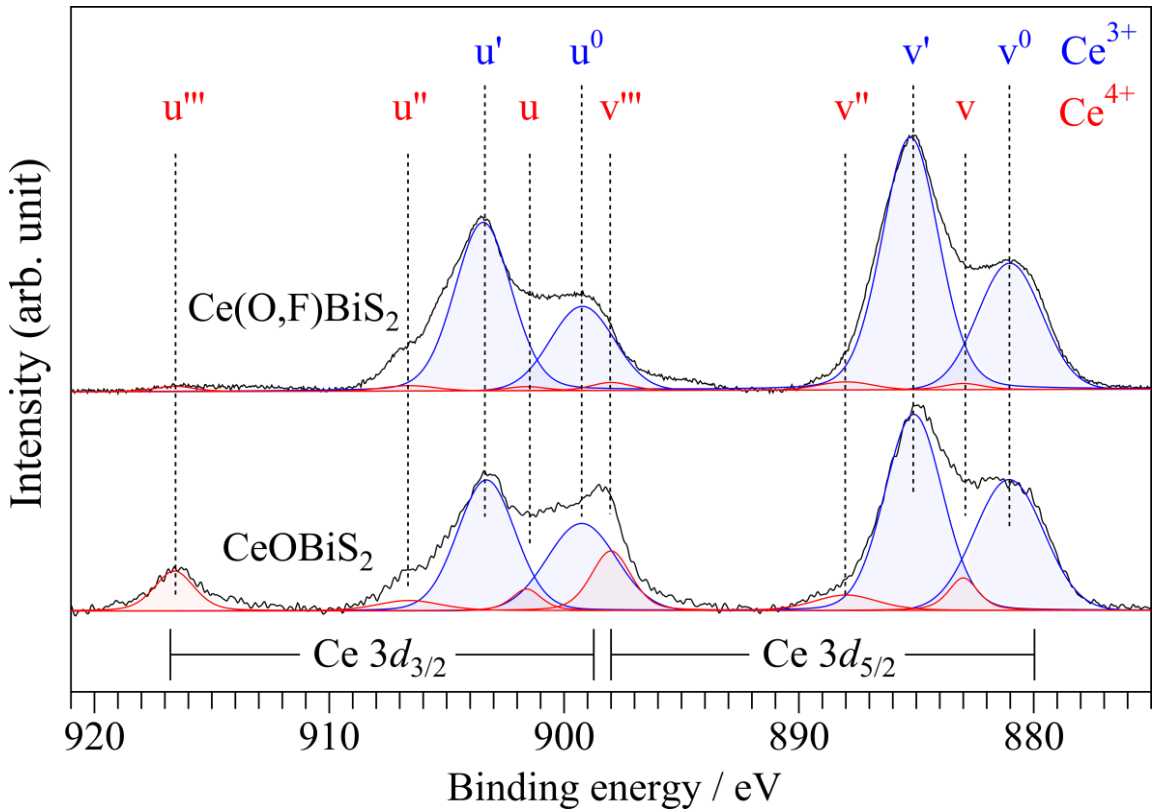

Figure 3. Ce $3d_{5/2,3/2}$ XPS spectra and the fitted curves for CeOBiS$_2$ (lower line) and Ce(O,F)BiS$_2$ (upper line) single crystals.

*3.3 Superconductivity in CeOBiS$_2$ and Ce(O,F)BiS$_2$ single crystals*

Figure 4 shows the temperature dependence of resistivity and magnetization for the single crystals of CeOBiS$_2$ and Ce(O,F)BiS$_2$. CeOBiS$_2$ clearly showed zero-resistivity around $T_c^{zero}$ ~1.3 K, suggesting that the "F-free" CeOBiS$_2$ shows superconductivity due to charge transfer from Ce valence fluctuation as speculated in Ref. [30].

The $T_c^{zero}$ of Ce(O,F)BiS$_2$ was ~3.1 K, and the corresponding Meissner signal was also observed in the magnetization measurement down until 2.0 K (inset of Fig.4). On the other hand, a magnetic anomaly around ~6 K has separately appeared, while the superconducting CeOBiS$_2$ did not show any magnetic behavior down to 2.0 K. Similar magnetic anomaly has been reported both in the single crystals and polycrystalline samples [2,5,16,17,19]. This magnetic anomaly has previously been reported as a coexistence of magnetic ordering and superconductivity [16,17]. However, the anomaly was also observed in SbS$_2$-based compound, Ce(O,F)SbS$_2$, which does not show superconductivity [41]. These facts imply that the magnetic anomaly is independent of the superconductivity, and it is caused by the common structure of Ce(O,F) layers, different from the superconducting BiS$_2$ layers. The appearance of anomaly in Ce(O,F)BiS$_2$ is attributed to originate from the valence modification in Ce atoms to Ce$^{3+}$-rich state by the F substitution.

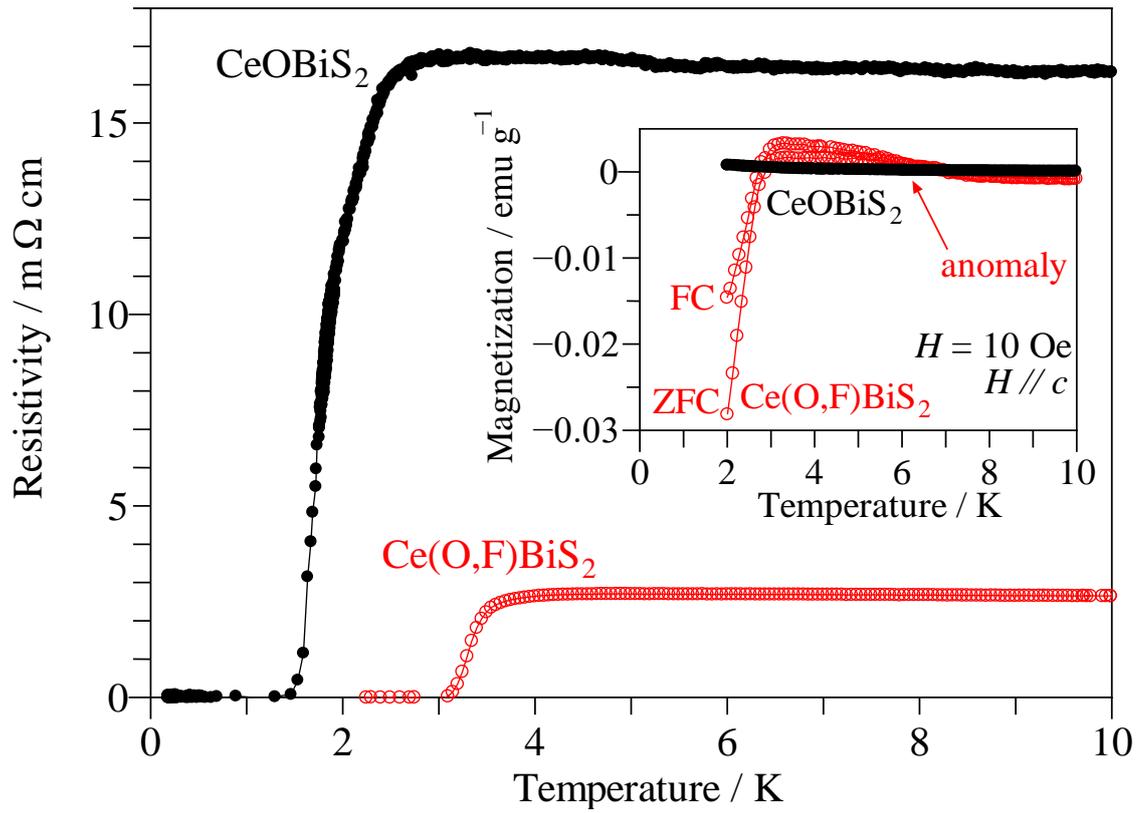

Figure 4. Temperature dependence of resistivity for the single crystals of $CeOBiS_2$ and $Ce(O,F)BiS_2$. The inset shows temperature dependence of magnetic susceptibility in zero-field cooling (ZFC) and field cooling (FC) mode. The magnetic field of 10 Oe was applied parallel to $c$-axis.

*3.4 High pressure effect on the superconductivity in CeOBiS$_2$*

$T_c$ in the BiS$_2$-based compounds is known as pressure sensitive [43,44]. The CeOBiS$_2$ is also expected to change its superconducting property under a high pressure of several GPa. Figure 5 shows the temperature dependence of resistivity under various pressure. The resistivity shows a broad hump-like behavior around ~100 K at ambient pressure. With increase of the applied pressure, the resistivity at room temperature tends to decrease. And the hump-like behavior was drastically suppressed above a pressure of 1.3 GPa, and then the resistivity starts to show metallic-like behavior. This tendency is similar to the result in the other BiS$_2$-based compound, EuFBiS$_2$, which also include a valence fluctuant Eu [45]. Although the mechanism of hump-like behavior and the reason why pressure can suppress it are unclear, the analogy is interesting.

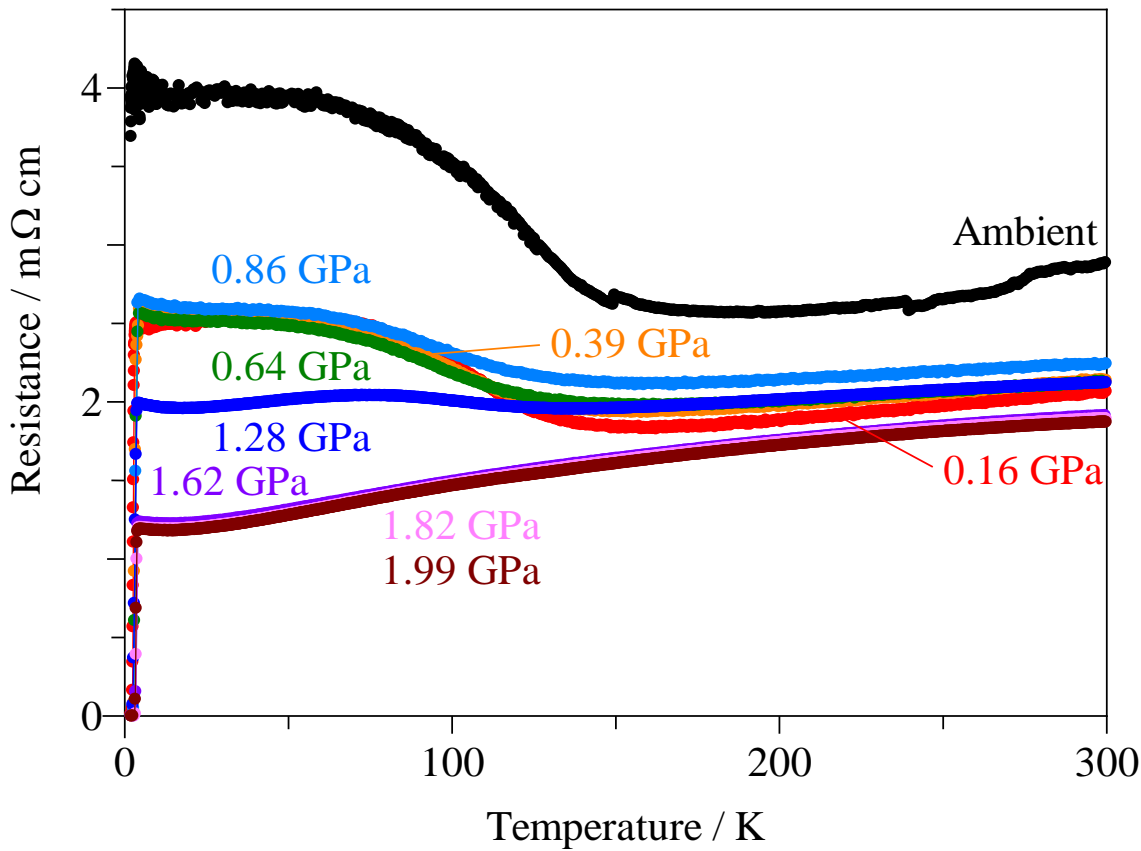

Figure 5. Temperature dependence of resistivity of CeOBiS$_2$ single crystals under various pressure.

Figure 6 shows an enlargement scale of the resistivity under pressure together with that was measured by ADR at ambient pressure (right hand side axis, the same data with Fig. 4). The onset $T_c$ ($T_c^{onset}$) was determined to the temperature at 95% of resistivity value at 10 K, and it was enhanced by applying pressure. It reaches to the maximum value of ~3.8 K, and zero-resistivity temperature ($T_c^{zero}$) was 3.2 K at a pressure of 1.62 GPa. After releasing the applied pressure, the $T_c$ returned back to the initial value of less than 2.0 K, and the large normal resistivity and hump-like behavior also resurfaced. Interestingly, the highest $T_c^{onset}$ and $T_c^{zero}$ are almost the same with those of F-doped single crystals as shown in Fig. 4. The carriers for superconductivity in CeOBiS$_2$ should be come from the mixed valence of Ce, because there is no dopant. Since the average valence of Ce was estimated to be Ce$^{3.19+}$ from XPS measurement, 0.19 electron carriers are excess from the charge neutrality in the blocking layers. Then it would be doped to the BiS$_2$ layers per Ce atom. On the other hand, the BiS$_2$ layer in F-doped crystal contains 0.27 electron carriers from substituted F of ~0.27 per CeO unit according to the EPMA estimation. These results suggest that the total carrier number is comparable to each other. There still remains a question why the $T_c$ of the CeOBiS$_2$ at ambient pressure is only 1.5 K, which is considerably lower than that of Ce(O,F)BiS$_2$. It might be attributed to the effective carrier numbers in BiS$_2$ plane. The BiS$_2$ layer in CeOBiS$_2$ is little bit corrugated compare to that is in Ce(O,F)BiS$_2$. When the external pressure was applied, the corrugation in the BiS$_2$ layers would be expected to be flatter than that is in ambient pressure. This results in increase of the effective carrier number contributed to the conduction nature. In other words, the carrier number in BiS$_2$ layers would be increased by applying pressure. It is reasonable that the comparable $T_c$'s of CeOBiS$_2$ under high pressure and Ce(O,F)BiS$_2$ are attributed to the total effective carrier numbers in the individual crystals.

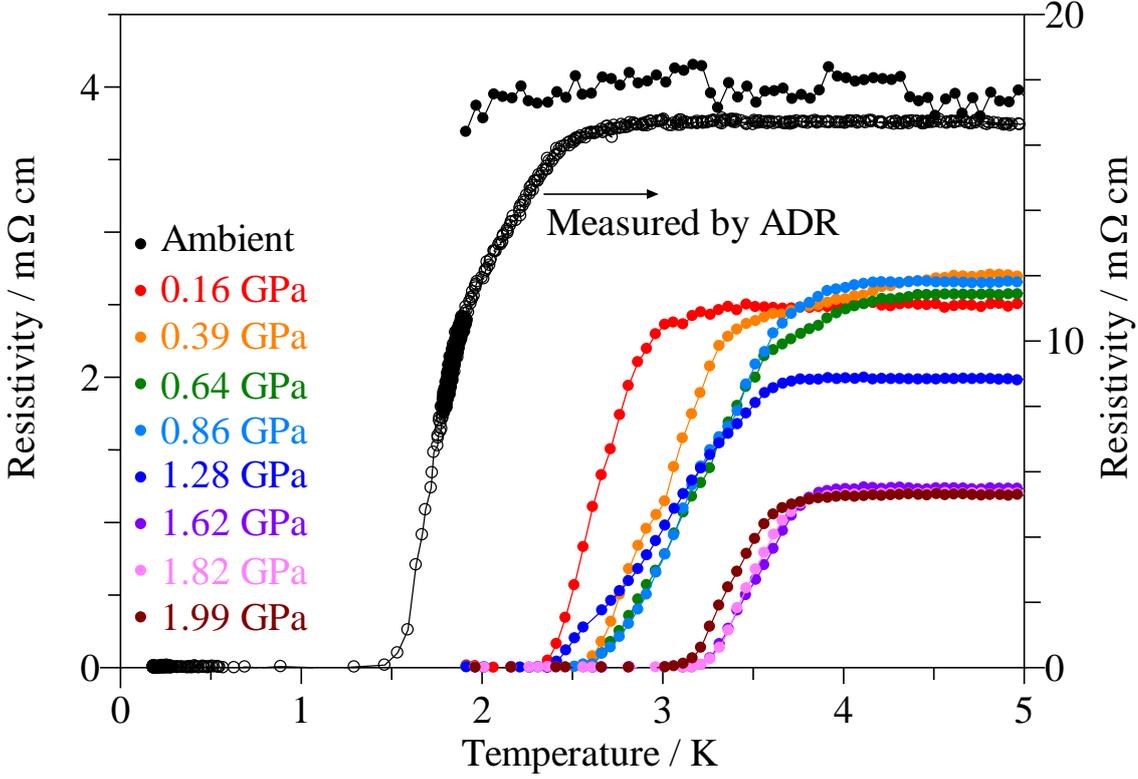

Figure 6. Enlarged scale of the temperature dependence of resistivity of CeOBiS$_2$ single crystals under various pressure. The right-hand side axis is used for the ambient one measured by using ADR option of PPMS (the same data with Fig.4).

The resistivity at 1.62 GPa was measured under various magnetic fields along the *c*-axis ($H//c$) and the *ab*-plane ($H//ab$) as shown in Figures 7(a) and 7(b), respectively. The superconductivity is significantly suppressed with increasing field in both directions. Figure 8 shows the temperature dependence of the upper critical field ($H_{c2}(T)$), which was plotted from the values of $T_c^{onset}$ in Fig. 7. The $H_{c2}(0)$ in $H//c$ and $H//ab$ were estimated to be $H_{c2}^{//c}(0)$ = 3240 Oe and $H_{c2}^{//ab}(0)$ = 7400 Oe, respectively, from the Werthamer-Helfand-Hohenberg (WHH) approximation for the Type II superconductor in a dirty limit [46]. The superconducting anisotropic parameter $\gamma = H_{c2}^{//ab}(0)/H_{c2}^{//c}(0)$ is determined to be 2.3. The $\gamma$ of Ce(O,F)BiS$_2$ was reported to 11-21 [26,47]. In case of La(O,F)BiS$_2$, the anisotropic parameter varies from 23 to 45 depending on its fluorine content [47]. $\gamma$ = 2.3 determined in this study is quite less anisotropic compare to that of the other BiS$_2$-based superconducting materials. The quite small anisotropy in the crystals might be reflected from the electronic structure. Further investigation is required.

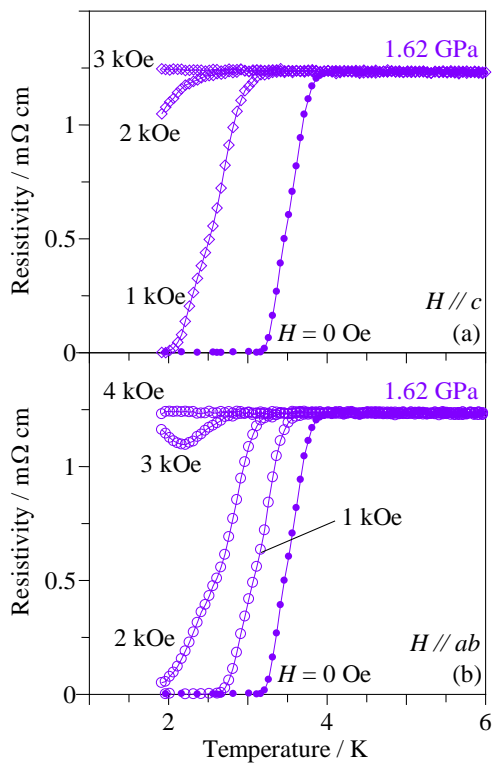

Figure 7. Temperature dependence of resistivity of CeOBiS$_2$ single crystals at a pressure of 1.62 GPa under various magnetic field in (a) $H//ab$, and (b) $H//c$.

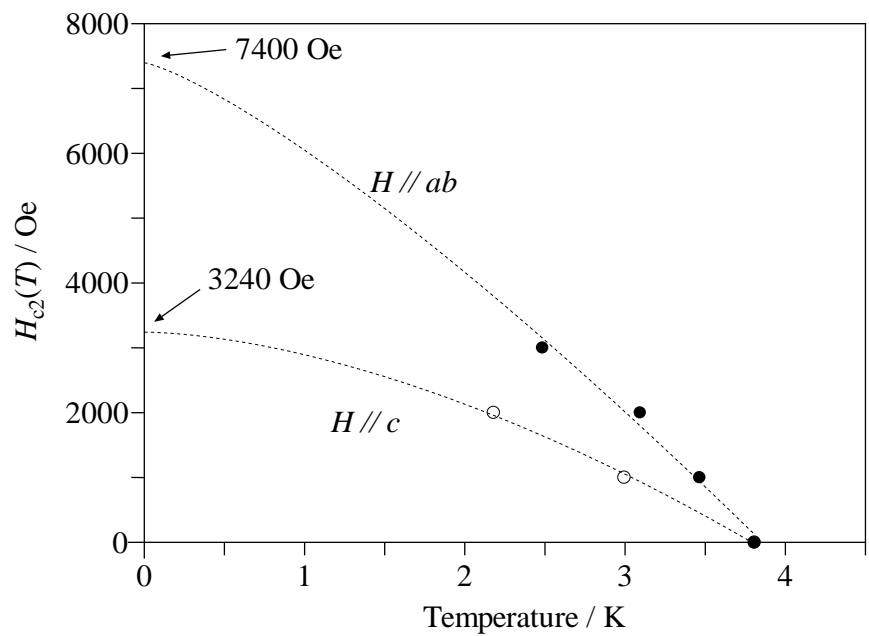

Figure 8. Temperature dependence of upper critical field $H_{c2}(T)$ of the single crystals of CeOBiS$_2$. The dotted line shows WHH approximation to estimate $H_{c2}(0)$.

## 4. Conclusion

F-free CeOBiS$_2$ single crystals were successfully grown using a high-purity CsCl flux. The crystal structure of CeOBiS$_2$ crystals was determined to have tetragonal crystal system with a space group *P4/nmm* (lattice parameters of *a* = 4.0189(6) Å, *c* = 13.573(2) Å). The BVS estimation and XPS analysis showed that the chemical state of Ce was in the mixed valence of Ce$^{3+}$ and Ce$^{4+}$. The single crystals show superconductivity with zero resistivity at ~1.3 K under ambient pressure. The onset temperature of superconducting transition was enhanced up to ~3.8 K by applying hydrostatic pressure. The superconducting anisotropic parameter was determined to be 2.3 from its upper critical magnetic field. The less anisotropic character in the compound will be expected to reveal by means of some electronic structural studies such as *ab-initio* calculations, and/or angle-resolved photoemission spectroscopy.

## Author Contributions

#M. Tanaka and M. Nagao contributed equally.

## Acknowledgements

The authors thank Dr. H. Okazaki (JASRI) and Prof. Dr. T. Yokoya (Okayama Univ.) for interpretation of XPS results, and also thank Dr. Y. Matsushita (NIMS) for his kind advices about structural analyses. This work was supported by JSPS KAKENHI [Grant No. 15K14113]; JST CREST [Grant No. JPMJCR16Q6].

## Appendix A. Supplementary data

The Supporting Information includes Crystallographic Information Files in CIF format for CeOBiS$_2$, Ce(O,F)BiS$_2$.